\documentclass{sigchi}

\toappear{Accepted for presentation at CSCW'15.}

\usepackage{balance}  
\usepackage{graphics} 
\usepackage{caption}
\usepackage{subcaption}
\usepackage{times}    
\usepackage{url}      
\usepackage{booktabs}

\makeatletter
\def\url@leostyle{%
  \@ifundefined{selectfont}{\def\UrlFont{\sf}}{\def\UrlFont{\small\bf\ttfamily}}}
\makeatother
\def\pprw{8.5in}
\def\pprh{11in}

\usepackage[pdftex]{hyperref}
\hypersetup{
pdftitle={SIGCHI Conference Proceedings Format},
pdfauthor={LaTeX},
pdfkeywords={SIGCHI, proceedings, archival format},
bookmarksnumbered,
pdfstartview={FitH},
colorlinks,
citecolor=black,
filecolor=black,
linkcolor=black,
urlcolor=black,
breaklinks=true,
}

\newcommand\tabhead[1]{\small\textbf{#1}}

\begin{document}

\title{MoodBar: Increasing new user retention\\ in Wikipedia through lightweight
socialization}


\numberofauthors{2}
\author{
  \alignauthor Giovanni Luca Ciampaglia\\
    \affaddr{School of Informatics and Computing,\\ Indiana University}\\
    \affaddr{919 E 10\textsuperscript{th} St., Bloomington, IN 47409, USA}\\
    \email{gciampag@indiana.edu}\\
  \alignauthor Dario Taraborelli\\
    \affaddr{Wikimedia Foundation, Inc.}\\
    \affaddr{149 New Montgomery St. Floor 6,\\ San Francisco, CA 94105, USA}\\
    \email{dtaraborelli@wikimedia.org}\\
}

\maketitle


\begin{abstract}

  Socialization in online communities allows existing members to welcome and
  recruit newcomers, introduce them to community norms and practices, and
  sustain their early participation. However, socializing newcomers does not
  come for free: in large communities, socialization can result in a significant
  workload for mentors and is hard to scale. In this study we present results
  from an experiment that measured the effect of a lightweight socialization
  tool on the activity and retention of newly registered users attempting to
  edit for the first time Wikipedia. Wikipedia is struggling with the retention
  of newcomers and our results indicate that a mechanism to elicit lightweight
  feedback and to provide early mentoring to newcomers improves their chances of
  becoming long-term contributors. 

\end{abstract}





\section{Introduction}

Improving the experience and retention of newcomers is one of the main
challenges that Wikipedia is facing these days. The base of contributors to the
``free encyclopedia that anyone can edit" has been suffering from a gradual
deterioration since reaching a tipping point at the end of a rapid growth phase.
The number of active editors on the English Wikipedia, defined as all registered
users who performed at least five edits in a given month, peaked around 2007 and
has been steadily decreasing ever since \cite{Suh2009, WSP2011}. This stagnation
in the contributor base primarily affects larger and more mature projects like
the flagship, English-language edition and less so smaller Wikipedia editions in
other languages. However, the trend remains a concern that the Wikimedia
Foundation -- the nonprofit organization that runs Wikipedia -- is currently
tackling with dedicated programs and interventions.

Long-term trends in new user retention have caused similar concerns. Longitudinal analysis of
cohorts of `new' editors -- editors who reach ten or more lifetime contributions
-- shows that one year after the 10-edit milestone the fraction of those who
performed one or more edits has gone down from about 40\%, for 2004 cohorts, to
about 10\% for those of 2009 \cite{WSP2011}. The phenomenon has been
acknowledged by the Wikipedia community and covered by popular press outlets,
which refer to it as the ``decline'' of Wikipedia \cite{Simonite2013,Economist2014}. 

Wikipedia is a complex socio-technical system, so various factors may contribute
to the decline in newcomer retention. In particular, the leading hypothesis
suggests that as a result of the desire to maintain high quality standards and
to fight vandalism \cite{Priedhorsky2007} over the years the Wikipedia editor
community has become impervious to new contributors \cite{Goldman2010}, who
nowadays have to cope with a daunting body of norms and policies
\cite{Beschastnikh2008} and sometimes unpleasant social exchanges, despite
producing the same rate of good faith contributions as users who joined the
project in earlier years \cite{Halfaker2013}. Such an unintended consequence is
not unlikely, since formal and informal norms about contributions
\cite{Ciampaglia2011}, socialization \cite{Choi2010}, and even language use
\cite{Danescu-Niculescu-Mizil2013} often calcify in online communities. Without
adequate support, newly recruited community members may have trouble conforming
to these tacit or explicit norms.

One possible solution to deal with a shrinking contributor base is to increase
the influx of new users, but competition for attention among different social
media platforms \cite{Backstrom2006, Ribeiro2014} implies that this process is
largely out of control of any single community -- Wikipedia included. A
complementary approach consists in increasing the participation and retention of
existing contributors once they join the project.

The effect of socialization -- the period during which a new member learns the
social norms and conventions of a group -- on retention has received much
attention in the social psychological literature on online communities
\cite{Burke2009,Choi2010,Farzan2012,Morgan2013}. Typical socialization tactics
include welcoming messages and the creation of safe sandboxing spaces where newcomers have an opportunity to learn. 
These efforts are usually tailored to small groups within a broader
community, and are often time-consuming for those who run them. Simpler
approaches may have the benefit of reaching a larger audience, but at the same
time their effect on retention and engagement may be limited.

\begin{figure*}[t]
  \centering
  \includegraphics[width=\columnwidth]{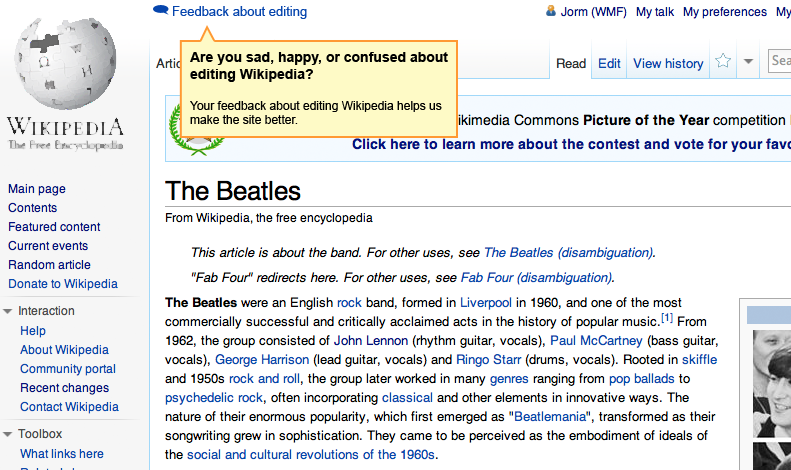}
  \hfill
  \includegraphics[width=\columnwidth]{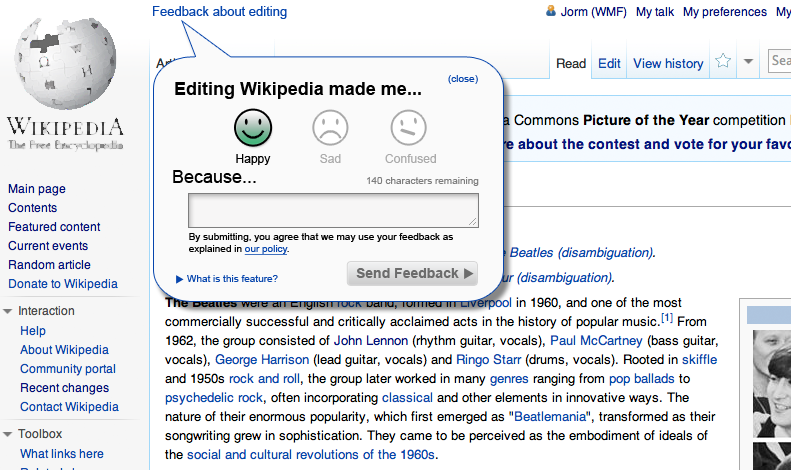}
  \caption{The MoodBar interface. Left: as soon the user clicks on the ``Edit''
    button of a page for the first time, a yellow tooltip appears inviting the user to send feedback about the editing experience. Right: as the user clicks
    on the MoodBar link, a larger window appears with a feedback form.} 
  \label{fig:MoodBar}
\end{figure*}

The goal of this study is to present results from an experiment on a simple,
lightweight socialization process and to understand its impact on the long-term
retention of newcomers in an open collaboration community. MoodBar is an
experimental feature introduced in Wikipedia between 2011 and 2013 with the goal
of eliciting feedback from newly registered users. It allows new users to send
feedback (or share their `mood') about their first editing experience on
Wikipedia (Figure~\ref{fig:MoodBar}). Feedback is posted on a public dashboard
and replied to by a team of experienced volunteers. As the name suggests, each
piece of feedback is characterized by a mood indicator (`sad', `happy' or
`confused') that gives a simple qualitative clue about the nature of the
message. Users experiencing issues with editing can report their problems,
typically via ``sad' or `confused' moods, and receive assistance to overcome
them. Users can also express gratitude towards the project as a whole or
happiness for completing a milestone, such as successfully saving a first edit.
Table~\ref{tab:examples} gives a few examples of feedback posted via MoodBar. 

\begin{table}
  \centering
  \footnotesize
  \begin{tabular*}{\columnwidth}{@{}p{\columnwidth}@{}}
    \toprule
    \multicolumn{1}{c}{\tabhead{Happy}} \\
    \midrule
    i loved it it was omg fun\\[.3em]
    I am a robot\\[.3em]
    Everything I have edited is correct.\\[.3em]
    I love editing but why do u delete my pages ?\\[.3em]
    I feel powerful, muahahaha.\\
    \midrule
    \multicolumn{1}{c}{\tabhead{Sad}} \\
    \midrule
    I can't edit protected articles\\[.3em]
    Wikipedia is a proganda. (\emph{sic.}) \\[.3em]
    I couldn't edit the intro. This needs a lot of work!\\[.3em]
    wikipedia doesnt let me troll :(\\[.3em]
    Site is slow again.\\
    \midrule
    \multicolumn{1}{c}{\tabhead{Confused}} \\
    \midrule
    i cant start a new page\\[.3em]
    too much code :[\\[.3em]
    The letters are hard to read\\[.3em]
    Can't change my user name\\[.3em]
    Editing in Arabic is somehow awkward because of RTL/LTR usual layout
    problems ...\\
    \bottomrule
    \end{tabular*}
    \caption{Examples of feedback posted via MoodBar.}
    \label{tab:examples}
  \end{table}

Our hypothesis is that, despite its extreme simplicity, lightweight socialization of the
kind provided by MoodBar can be effective at improving the chances that a newly
registered user survives as a long-term contributor. By using a combination of observational and
experimental methods, we aim to address the following research
questions:

\textsc{rq 1}. \emph{Do users post feedback about their \emph{early} editing
experience, or is feedback posted at a later stage?}

\textsc{rq 2}. \emph{Is reporting feedback associated with a higher productivity by newcomers in the short term?}

\textsc{rq 3}. \emph{Does posting feedback and receiving a response improve long-term retention of newcomers?}


In the first part of the study we characterize how MoodBar is used and by whom.
We analyze how the tool is utilized with respect to the designers'
intentions, that is, to report feedback about early editing experience on
Wikipedia. We then focus on the productivity of those who used MoodBar to report
feedback, and compare them to the larger population of newly registered users. We find
that in the early stage of activity MoodBar users are in general more productive
than users who do not share their mood and that productivity increases for those users who received a response to their feedback. Usage of MoodBar appears
to be strongly associated with higher levels of contribution.

A potential explanation of this result is a self-selection bias. Sending
feedback requires locating a small link at top of a page and writing a short
(140 characters) message. Having the intention and the ability to send feedback
may itself be a strong indicator of the presence of cognitive and social skills
required to succeed in a complex socio-technical environment such as Wikipedia
\cite{Egan1988}. A purely observational method can only reveal whether usage of
MoodBar is associated with higher rates of activity and retention, not that the
association is a form of causation.

In the second part of this study, we analyze the long-term retention of a large
sample of newly registered Wikipedia users. We analyzed the behavior of a group
of users who were not able to see or post messages via MoodBar, because the link
to post a mood message was suppressed. This group of users served as our control
group. We compared their long-term retention to that of a group of users who had
regular access to MoodBar (the treatment group), registered shortly before users
in the control group.

We found that after 180 days since registration users in the treatment group had
a small but significant increase in retention compared to users in the control
group. Significant differences in retention between the two groups emerge as
early as 120 days after registration. Since the only known difference between
the two groups is the availability of MoodBar, this rules out the presence of
selection bias and suggests that MoodBar has a positive effect on long-term
retention of these users. It should be noted that only a small fraction of newly
registered users has the chance of interacting with MoodBar, thus while at the
group level the overall effect of MoodBar is very small ($d = 0.22\%$), we
estimate that the relative increase must be higher. 

The rest of the paper is organized as follows: in the next section we review related work on early socialization in online
communities. We then present the methodology used in the study: we describe
how MoodBar works, how the data used in this study was collected, and present 
the methodology used in the remainder of this work. We conclude by discussing design 
implications of our findings in the final section of the paper.

\section{Related work}

A large literature has studied incentives and drivers of participation in online
communities, with a focus on early socialization. Early research on Wikipedia
and open source software projects suggests that a mix of intrinsic motivation
and extrinsic rewards drives participation \cite{Hars2001, Lerner2002,
Zhang2006}. Top contributors may have strong intrinsic motives to participate
\cite{Panciera2009}. Non-monetary rewards such as acknowledgements
\cite{Beenen2004, Ling2005, Rashid2006, Cheshire2008}, badges \cite{Restivo2012,
Anderson2013}, and gamified feedback \cite{Deterding2011} have been shown to increase
engagement of users. Certain forms of reward can exert fine-grained control,
even though instilling long-term behavior still proves to be difficult
\cite{Anderson2013}.

Besides individual incentives, previous studies also stressed the
importance of the initial period of socialization in online groups. A successful
early socialization experience is associated with, and sometimes even predicts,
increased engagement in mailing lists \cite{Backstrom2007}, newsgroups
\cite{Joyce2006}, social networks \cite{Burke2009}, and Wikipedia
\cite{Choi2010, Morgan2013}, to cite a few. However, the causal structure
between socialization, motivation, and participation is still not entirely
clear. Strong motivational factors, perhaps in conjunction with individual-level
skills \cite{Egan1988}, may be the cause for both a successful early
socialization stage and a later long-term participation. To further establish a
causal connection, controlled and field experiments on groups of limited size
have been performed, with encouraging results: sharing in a digital information
good is increased by social incentives \cite{Cheshire2007}, personal messages
improve the retention of newcomers to Wikipedia who had their edits rejected
\cite{Geiger2012}, and top contributors in a Q\&A community contributed more on
the long term if they had received a personalized socialization experience
\cite{Farzan2012}. 

\section{Methods}

MoodBar is an experimental Mediawiki extension, loosely inspired by the Mozilla
Firefox Input system \cite{Mozilla2014}. It allows newly registered users to
send their feedback (or share their `mood') about their first edit experience on
Wikipedia (Figure~\ref{fig:MoodBar}). Because it is meant to
elicit feedback from newly registered users at their first edit, MoodBar activates
itself only after the user attempts to edit a page for the first time. Upon
clicking on the `Edit' button, a link appears in the upper left area of the
screen together with a brief notification in the form of a tooltip,
(Figure~\ref{fig:MoodBar}, left).

\begin{figure}[t]
  \centering
  \includegraphics[width=\columnwidth]{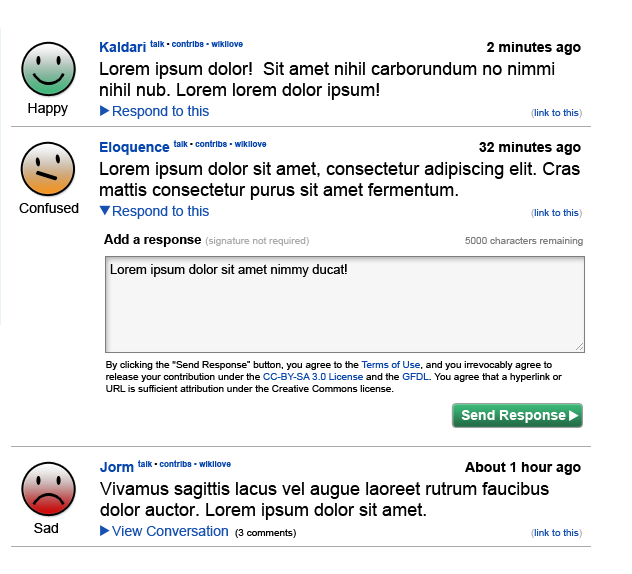}
  \caption{A mockup of the Feedback Dashboard displaying mood messages posted by newcomers that volunteers can respond to.}
  \label{fig:Feedback-Dashboard-Phase2-AddResponse-Step2}
\end{figure}

Clicking on the MoodBar link a non-modal window appears
(Figure~\ref{fig:MoodBar}, right), allowing the user to select one out of three
moods and post a short feedback message. Feeback posted by the user is displayed
on a public Feedback Dashboard
(Figure~\ref{fig:Feedback-Dashboard-Phase2-AddResponse-Step2}), where it can be
processed by a team of experienced volunteers. When MoodBar feedback is replied
to by a volunteer, a message is automatically published on the talk page of the
original poster, and if the user has a verified email address on file an email
notification is sent. The original poster can then read the response and, if
they find it useful, publicly mark it as such on the Feedback Dashboard.

\subsection{Datasets}

The data we used in this study was extracted from the database of the English
Wikipedia, where MoodBar was deployed between July 2011 and February 2013. An
earlier version of MoodBar allowed users to report a mood without inserting any
message and did not not show any tooltip notification upon activation. In a
second iteration, finally deployed in December 2011, the text message was made
mandatory and a prominent tooltip was introduced. In this study we only consider
data collected from this second version of MoodBar. The samples of users we
considered for our analysis are summarized in Table~\ref{tab:samples}.

\begin{table}[t]
  \centering
  \begin{tabular*}{\columnwidth}{@{}lccr@{}}
    \toprule
    \tabhead{Group} & \tabhead{Start} & \tabhead{End} & \tabhead{$N$} \\ 
    \midrule 
    Historical  & 2011-12-14 & 2012-05-22 & $528,891$\\ 
    \midrule
    Hist. (Reference) & 2011-12-14 & 2012-05-22 & $515,438$\\
    Hist. (Feedback)  & 2011-12-14 & 2012-05-22 & $8,599$\\ 
    Hist. (Feed.+Resp.) & 2011-12-14 & 2012-05-22 & $4,164$\\ 
    Hist. (Feed.+Useful) & 2011-12-14 & 2012-05-22 & $690$\\ 
    \midrule
    Treatment   & 2012-05-23 & 2012-06-14 & $64,652$\\
    Control     & 2012-06-15 & 2012-06-29 & $40,389$\\ 
    \bottomrule
  \end{tabular*}
  \caption{The samples of users of this study. Users who registered an accounts
    on the English Wikipedia and clicked on the `Edit' button at least once are
    assigned to a group based on the registration date.}
  \label{tab:samples} 
\end{table}

\subsection{Observational study}

We use the `Historical' sample (Table~\ref{tab:samples}, top) to perform the
observational part of the study. This sample spans about 5 months of new account
registrations. Automated accounts (bots) were filtered out using a list of such
accounts. We compute two metrics for users in this sample. The first metric is
the time lag between the activation of MoodBar (i.e. the first time the `Edit'
button is clicked) and the first feedback posted by a user, if any. If a user
did not send any feedback, we include a censored observation, indicating that
she could still potentially send one in the future. To estimate the survival
rate and filter inactive accounts, we considered only users who performed at
least one edit to compute this metric. This reduced the sample size to
$N=95,586$ users, of which only $19,219$ sent feedback (i.e. actual, uncensored
observations). The time stamp of the first `Edit' click was tracked by the
`EditPageTracking' extension \cite{EditPageTracking2014}, an ancillary extension
of MoodBar. 

The second metric, which we use to quantify short-term productivity, is the
cumulative number of contributions measured at 1, 2, 5, 10, and 30 days of days since
registration. We counted contributions to any type of page on Wikipedia,
including project pages and user talk pages. Contribution data was collected via
the `UserDailyContribs' extension \cite{UserDailyContribs2014}. We computed
these metrics for the full `Historical' sample, but we further distinguish users
based on their type of interaction with MoodBar (Table~\ref{tab:samples},
middle). We consider four mutually exclusive groups: all users who attempted to
edit and saw the tooltip but did not send any feedback (`Reference'), those who
did posted feedback but received no response (`Feedback'), those who posted
feedback and got a response but did not mark it as helpful (`Feed.+Resp.'), and
those who did mark the response as helpful (`Feed.+Useful'). In the rare case
that a user sent multiple feedback messages, we use only the first feedback to determine
which of the three sub-groups she belongs to.

\subsection{Natural experiment}

In the second part of the study, we test the socialization effects of MoodBar on
long-term retention by identifying a treatment and control group
(Table~\ref{tab:samples}, bottom). 

The metric we use to quantify long-term user retention is the survival
probability at $t = 180$ days since registration, which is defined as the
fraction of users who made at least one contribution at any time $t' > t$. To
compute it, we collected user contribution data from both cohorts up to a year
after the end of the registration window of the latest of the two cohorts. This
amounts to an observation window for $t'$ of at least 27 weeks.

To identify the control group we suppressed the link to MoodBar for all users
registered during a specific time window. Other approaches, for example showing
the link to only a subset of users during the same period could have been
possible too, but since we needed to run the experiment on Wikipedia's
production servers, we opted for the simplest to implement and deploy.

\subsubsection{Power analysis}

Users who sent a feedback with MoodBar are a tiny fraction of the overall
population of users who saw the MoodBar link. A conservative upper bound from
the `Historical' sample taking into account only users who clicked at least
once on the `Edit' button is equal to $2.5\%$. 


This means that when diluted within the broader population of registered users the
difference in retention between the treatment and the control group, if any, is
going to be very small. We performed \emph{a priori} power analysis to determine
the minimum sample size before collecting data for the treatment and control
group cohorts and used the `Historical' sample to do so, as the most
recent available sample of users (see Appendix). 


The power of a statistical test is the probability of correctly rejecting the
null hypothesis. The minimum sample size required to detect a difference in
retention equal to $\Delta$ with a test of power $1 - \beta = 80\%$ and
significance $\alpha = 5\%$ is given by:
\begin{equation} N^\ast = \frac{16 \sigma^2}{\Delta^2} \label{eq:samplesize}
\end{equation}

In our case, using the first estimate of $\pi_m$ (see Appendix), we obtain
$N^\ast = 130,763$ users. Assuming that an underestimate to the average daily of
new registered users who activate MoodBar is about $1,600$, this would require
us to keep MoodBar disabled for at least $82$ days in order to gather enough
users for the control group cohort. 




\subsubsection{UI manipulation}

Instead of opting for such a long window, we decided to (1) elicit more feedback by
increasing the saliency of MoodBar, and (2) increase the rate of replies on the
Feedback Dashboard. Receiving a reply to the feedback is associated to an even
higher value of $\pi_m$, and thus of $\Delta$. Since we expect the impact of
MoodBar on retention to be caused by socialization, i.e. by feedback responses,
we calculated that a conservative boosting factor of $b = 1.5$ in the fraction of
users who received a reply to their feedback would result in $N^\ast = 15,329$
and in a registration window of at least $10$ days. Based on these
considerations, we chose a registration window for the control group of two
weeks. 

In order to improve the saliency of MoodBar we manipulated the user interface
and increased the size of the MoodBar link, and allowed the notification tooltip
to stay on screen for a few more seconds before disappearing (see
Figure~\ref{fig:MoodBar}). To increase the rate of replies we issued a call to
actions to the team of volunteers by asking them to provide replies to incoming
posts via the Feedback Dashboard. 


\section{Results}

\subsection{Observational study}

\begin{figure}[ht]
  \centering
  \includegraphics[width=\columnwidth]{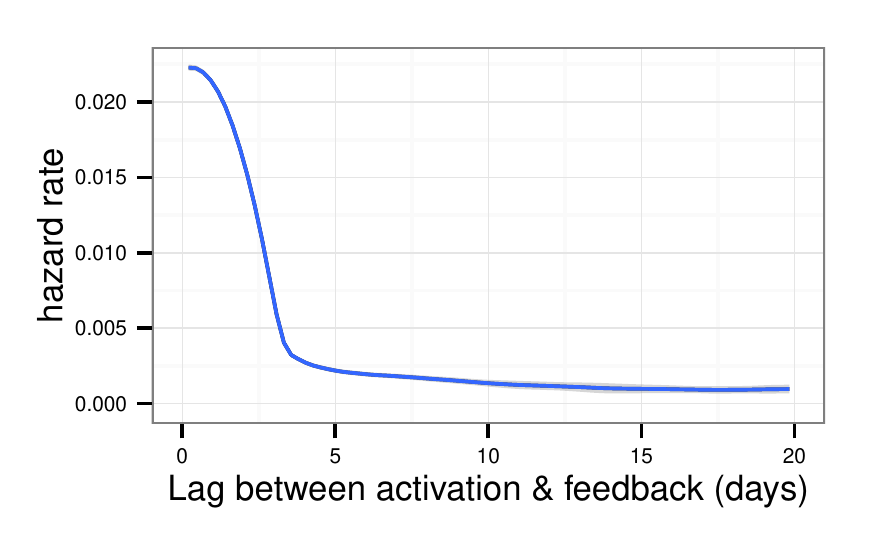}
  \caption{Smoothed hazard rate (see text for definition) of the first MoodBar
    feedback in the 20 days since the earliest click on the `edit' button.
  Resampled 95\% confidence bounds are smaller than the width of the line.}
  \label{fig:hazard} 
\end{figure}

Our first question is whether MoodBar is effectively being used to post feedback
about \emph{early} editing experience. To answer this question we look at the
hazard rate, the probability of sending a feedback at a specific time lag given
that the user has not sent any feedback before. Figure~\ref{fig:hazard} shows
the hazard rate curve estimated on the `Historical' sample. The time lag on the
$x$-axis is computed since the activation of MoodBar, i.e. since the time the
user clicks on the `Edit' button for the first time. The hazard drops after the
first few days: on the tenth day the hazard rate is about 14 times lower than on
the first day. This confirms our hypothesis (\textsc{rq 1}) that MoodBar is
indeed used to report on early editing experiences. Looking at the actual mood
reported, we find that the median time to report `confused' and `happy' moods is
roughly 30 minutes, while `sad' moods take longer, about 2 hours.

Our second question concerns the productivity of MoodBar users against the
broader population of registered Wikipedia users who attempted to edit
(`Reference'). Table~\ref{tab:overdispersion} shows the average cumulative
number of contribution of users in the `Historical' sample at various days since
registration. MoodBar users are more productive than users in the `Reference'
group. At 30 days since registration those who received a useful feedback
(`Feed.+Useful') where about 9 times more productive than those who never sent a
feedback. 

\begin{table*}[t]
  \centering
  \begin{tabular}{@{}cr@{.}lr@{.}lr@{.}lr@{.}lr@{.}lr@{.}lr@{.}lr@{.}l@{}}
    \toprule
    \tabhead{Days since registration}  & \multicolumn{4}{c}{\tabhead{Reference}} & \multicolumn{4}{c}{\tabhead{Feedback}} & \multicolumn{4}{c}{\tabhead{Feed.+Resp.}} & \multicolumn{4}{c}{\tabhead{Feed.+Useful}}  \\
    \midrule
    1 & 1&83 & (4&95) &  3&90 & (9&34) & 3&70 & (8&47) & 6&82 & (16&96) \\
    2 & 1&95 & (5&63) &  4&52 & (12&45) & 4&18 & (10&32) & 8&09 & (19&09) \\
    5 & 2&20 & (7&56) &  5&83 & (19&41) & 5&22 & (15&44) & 11&38 & (26&65) \\
    10 & 2&51 & (11&10) &  7&54 & (27&10) & 6&37 & (22&50) & 16&27 & (46&10) \\
    30 & 3&16 & (24&20) &  11&72 & (53&50) & 9&55 & (58&20) & 27&75 & (90&70) \\
    \bottomrule 
  \end{tabular}
  \caption{Mean and standard deviation (in parentheses) of the number of
    contributions at 1, 2, 5, 10, and 30 days since registration for the
    different sub-samples of the `Historical' group.} 
  \label{tab:overdispersion}
\end{table*}

We confirmed the increased productivity of MoodBar users by performing a
regression analysis in which we control for the month of registration, to
account for seasonality, and the lag between account registration and activation
of MoodBar. Regarding the choice of the regression model, it should be noted
that the data is overdispersed, with a variance-to-mean ratio $\sigma^2 / \mu >
1$ in all cases. To address this issue we used a negative binomial generalized
linear model. The ratio of residual deviance to degrees of freedom in the fitted
model is very close to $1$ ($D / d.f. = 0.9478$), which indicates a good fit. 

We also checked the plausibility of a self-selection bias via a simple
observation. By definition, since their feedback did not receive any reply,
users in the `Feedback' group missed the opportunity for socialization offered
by MoodBar, so these users should not be different than the reference group.
However, according to the regression model, these users are significantly more
productive than the reference group by a factor of $2.36$ times ($p < 0.001$).
We take this as strong evidence for the existence of self-selection bias among
MoodBar users (\textsc{rq 2}).


\subsection{Natural experiment}

\begin{figure}[h]
  \centering
  \includegraphics[width=\columnwidth]{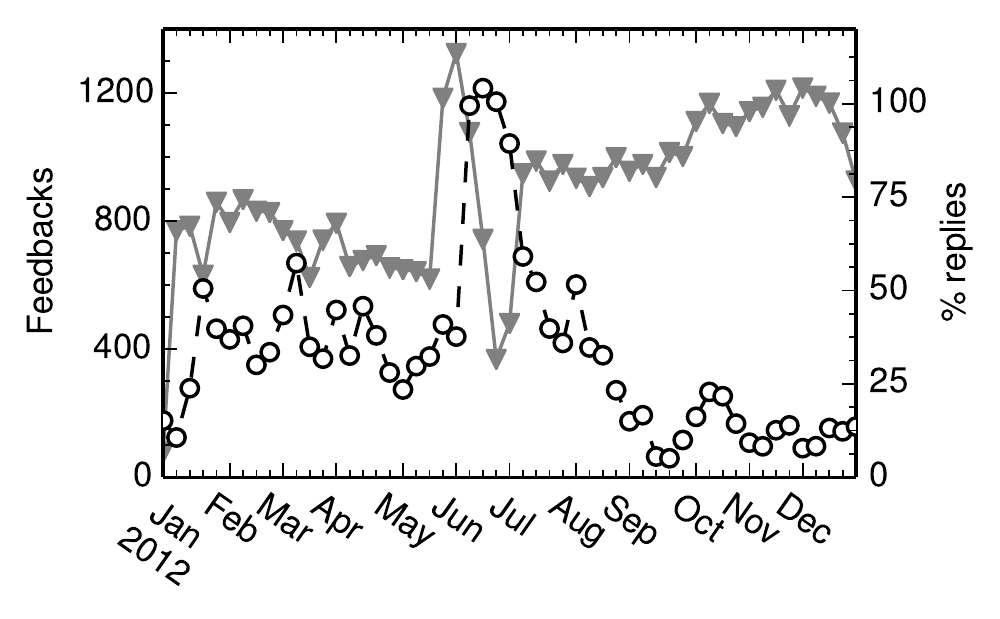}
  \caption{Weekly volume of MoodBar feedback messages posted (gray triangles) and weekly ratio
  of replies to feedback messages (white circles).}
  \label{fig:feedback_lineplot}
\end{figure}

We assessed the impact of the increased saliency of the MoodBar link by
measuring the weekly volume of feedback messages posted and the weekly ratio of
replies to feedback posted during the first seven months of 2012
(Figure~\ref{fig:feedback_lineplot}). Feedback volume spiked during the
registration window of the treatment group, and dropped on the third week of
June as a consequence of the suppression of the MoodBar link. The volume did not
go to zero during this blackout, as MoodBar was still available to users
registered before the experiment. For the same reason, the spike in feedback
volume reflects feedback posted both by newly and previously registered users. 

We can estimate the actual boost factor by comparing the volume of
feedback before the \textsc{ui} tweak to that after the end of the control group
window. This method yields a boosting factor of $b = 1.54$. The weekly ratio of
replies to feedback increased as well, reaching a peak two weeks after the
beginning of the registration window of the control group. On that week, the
replies ratio was slightly above $100\%$, since volunteers replied to older
feedback too. 

%

\begin{figure*}[t]
  \centering
    \includegraphics[width=\columnwidth]{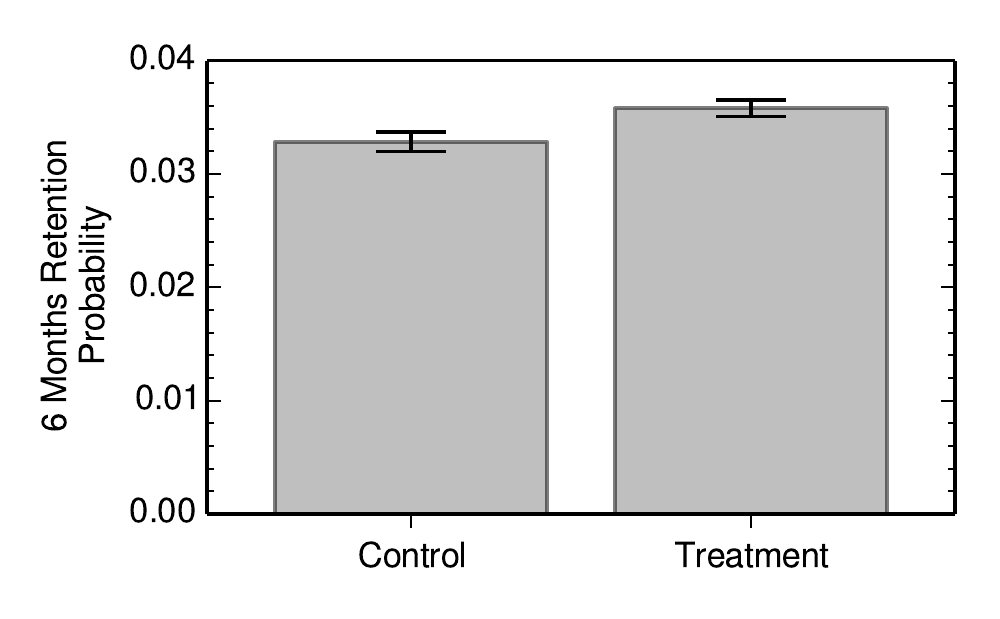}
  \hfill
    \includegraphics[width=\columnwidth]{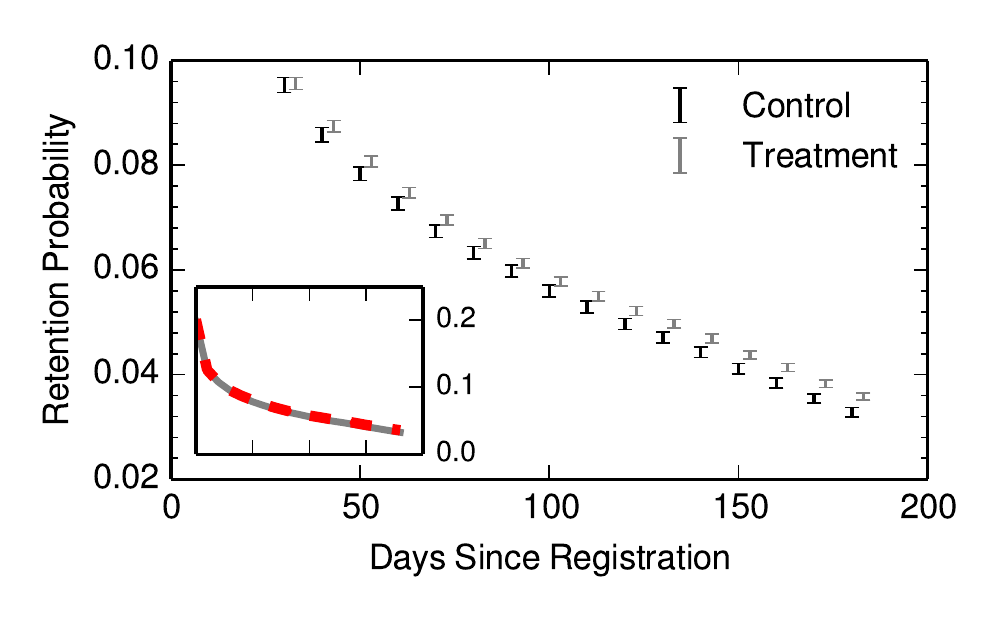}
    \caption{\emph{Left}: Retention probability at 180 days after registration
    for the two experimental groups. Error bars are 95\% confidence intervals.
    \emph{Right}: A zoomed detail of the full retention curves of the two
    groups. For ease of comparison, the treatment curve is shifted on the right
    of 3 days. Error bars are 95\% confidence intervals. The inset shows the
    full curve with no shifting. The red dashed line is the treatment group.
  } 
  \label{fig:retention}
\end{figure*}

To answer the last research question -- whether or not MoodBar has a positive
effect on long-term retention (\textsc{rq 3}) -- we seek to reject the null
hypothesis that the proportion of retained users at 180 days in the control and
treatment groups is the same, that is, $H_0 : \pi_m = \pi_{\not m}$. The two
samples are independent, so we use a two-tailed test based on the normal
approximation and a pooled estimate of the variance. Using this test, we were
able to reject $H_0$ ($z = -2.56, p = 0.01$). Figure~\ref{fig:retention} (left)
plots the estimates computed from the data, showing a small but clear
difference. \emph{Post hoc} power analysis yields a statistical power of $1 -
\hat\beta = 73.88\%$, consistent to our original expectations.

We also looked at the evolution of differences in retention during the whole 6
month period. Figure~\ref{fig:retention} (right) shows the full
retention curves in the interval $\left[ 0, 180 \right]$ days. Differences in
retention between the two groups, as evidenced by the overlap between the
respective 95\% confidence intervals, emerge for a brief period around 50 days.
From 120 days onward the two intervals no longer overlap.

\section{Discussion}

The two groups we considered in the experiment were cohorts of users registered
in two consecutive windows lasting about a month in total. Because we could not
control who registered during both periods, our methodology is closer to the
category of natural experiments than to that of controlled randomized designs.
To the best of our knowledge, no other treatment was administered to users in
these cohorts during that period and given the short time frame considered we
make the assumption that the composition of the two groups in terms of
individual-level skills is homogeneous. Therefore, we can reasonably assume that
only difference between the group that is driving the effect is access to
MoodBar for users in the treatment group.

One counterintuitive aspect of our experiment is that despite its robustness,
the group-level effect of MoodBar on retention, being diluted on the whole
sample of registered users, is small in absolute terms. This is a consequence of
the limited saliency of MoodBar within the \textsc{ui} of Wikipedia, which
results in a very small value of $p$, the probability that an editor sends a
feedback. This means that very few people used MoodBar, and thus their impact on
the overall group retention is limited.

Things change dramatically at the individual level, where we see some evidence
that MoodBar has marked effects on retention and engagement. The relative
increase of the retention of users with successful socialization experience
(`Feed.+Useful') over the baseline of those who did not send any feedback
(`Reference') is $707\%$. To at least partially account for the self-selection
bias we can use as baseline those who posted feedback but did not receive any
reply (`Feedback'). These users never enjoyed the benefits of socialization, so
their higher retention rates must be only due to their individual
skills/motivation. The relative increase in this case is a more conservative
$143\%$. This line of reasoning, however, does not completely rule out
self-selection bias, so these figures should be taken at face value and not as
estimates of the individual-level retention increase due to MoodBar. 

Finally, the engagement increase of different sub-groups of MoodBar users in the
`Historical' sample (see Table \ref{tab:overdispersion}) suggests that most
benefits of using MoodBar come from receiving a useful reply, which confirms the
idea that newcomers benefit the most from active socialization exchanges with
existing users, and not just from simply posting a feedback.

\section{Conclusion and future directions}

Our findings provide evidence that early mentoring of newcomers in an open
collaboration system through lightweight socialization tools such as MoodBar
improves their engagement and retention.

We found that eliciting feedback via simple \textsc{ui} manipulations is an
effective way to reach users who are at the earliest stages of their editing
experience and might otherwise be unable to receive mentoring and support
(\textsc{rq 1}). We also found that these feedback mechanisms tend to
self-select users who have a higher natural propensity to become active
contributors (\textsc{rq 2}). Finally, we found evidence that early interaction
with these feedback mechanisms has a significant, long-lasting effect on the
retention of these users (\textsc{rq 3}).

Considering that experienced contributors perform a large amount of work on
Wikipedia, we submit that designing interfaces like MoodBar could help mitigate
the stagnation and newcomer retention problem Wikipedia is currently facing.

There are a number of limitations and possible research directions that this
study did not explore and future research should address.

Despite the fact that a relative small number of experienced users in the
``response team'' successfully managed to work through a backlog of feedback
messages to respond to, our results do not indicate how scalable this approach
would be and at what point the ability to socialize a substantially larger
number of newcomers would start to break down. We provided evidence that
lightweight interaction, based on very short messages and responses, can go a
long way in socializing new users but also indicated that the workload was
manageable at the current scale. Research indicating that canned, depersonalized
messages can negatively impact newcomer retention suggests that any attempt to
run the MoodBar model at a larger scale would need to assess the risk of
depersonalized communication \cite{Geiger2012}.

We did not perform any kind of qualitative analysis on the type of messages
elicited by newcomers to try and understand how self-reported mood and the
specific issues being discussed may be associated with engagement and retention.
As a result, we do not know what socialization strategy is the most effective,
and what aspect in the socialization process afforded by MoodBar drives editor
retention. Qualitative analysis of messages exchanged in the context of this
lightweight process should be compared with findings from previous studies where
more in-depth socialization strategies were considered \cite{Morgan2013}. 

Finally, in this study we only considered newcomers on the English Wikipedia.
Data from other Wikipedia language editions indicates that other communities
have different retention rates for newcomers. As a result, our findings may not
immediately generalize to other Wikipedia communities governed by different
norms or practices, or composed by a substantially different user demographics.
At a broader level, however, our finding applies to any online community where
users contribute content, such as ratings, reviews, or photos; wikis are just an
example of open collaboration communities and any of these has its own
set of norms about contribution. In fact, the problem of socializing a suddenly
growing number of newcomers, or ``eternal September'', dates back to the early
period of \textsc{usenet} groups. Our results thus provide evidence that
lightweight socialization tools could make newcomer socialization sustainable in
other online communities too.

While MoodBar was retired as an experiment from the English Wikipedia in 2013,
it is still in use in other Wikimedia projects, so the methodology used in this
study could be replicated to other projects to provide an additional validation.

\section{Acknowledgments}

The authors would like to thank all Wikipedia contributors who volunteered at
the Feedback Dashboard during the course of the experiment, Jonathan Morgan for
useful comments, Howie Fung, Benny Situ, and the Editor Engagement Team at the
Wikimedia Foundation for their assistance with setting up the experiment, and
the Wikimedia Foundation for supporting the work. Giovanni Luca Ciampaglia was
partially supported by the Swiss National Science Foundation under fellowship
no. 142353 and by the Center for Complex Networks and Systems Research at
Indiana University.

\appendix

\section{Power Analysis}

Since we had a sample where any user could potentially access MoodBar, we
estimated the expected difference in retention between treatment and control
from that of the whole sample by subtracting from it the contribution due to
MoodBar users. Let us denote with $p$ the probability that a user posts
feedback using MoodBar, with $\pi$ the probability of retention of a user, with
$\pi_m$ the probability of retention of a MoodBar user, and with $\pi_{\not m}$
the probability of retention of a non-MoodBar user (i.e. a user who does not
send any feedback). We are interested in estimating the difference $\Delta =
\pi_m - \pi_{\not m}$. It is the case that:
\begin{equation} \pi = p \pi_m + \left( 1 - p \right) \pi_{\not m} \label{eq:pi}
\end{equation}

And so, substituting for $\pi_{\not m}$ in $\Delta$ we have:
\begin{equation}
  \Delta = \frac{p}{1 - p}\left( \pi - \pi_m \right)
  \label{eq:Delta}
\end{equation}

\noindent which we can easily estimate from the `Historical' sample. We quantify
the effect size using the standardized difference, or Cohen's $d = \Delta /
\sigma$ \cite{Wilkinson1999}, where $\sigma$ is the pooled standard deviation of
the two groups:
\begin{equation} \sigma = \sqrt{\frac{\left( N_m - 1 \right) \sigma_m^2 + \left(
    N_{\not m} - 1 \right) \sigma_{\not m}^2}{N_m + N_{\not m} - 2}}
    \label{eq:sigma}
\end{equation}

\noindent where $N_m$ and $N_{\not m}$ are the number of users who sent at least
a feedback and of those who did not send any feedback, respectively, and
$\sigma_m$ and $\sigma_{\not m}$ the standard deviations. On the `Historical'
sample, $d = 0.22\%$ for the retention at $t = 30$ days since registration. If
we compute $\pi_m$ and $p$ on the subset of MoodBar users who received a reply
to their feedback, $d = 0.09\%$.

%
%
%
%
%
\balance

\bibliographystyle{acm-sigchi} \bibliography{cscw_moodbar}
\end{document}